\documentstyle[twocolumn,aps]{revtex}

\input epsf

\newcommand{\vecc}[1]{\mbox{\boldmath $#1$}}
\newcommand{\dd}{\mbox{d}}

\newcommand{\ii}{\mbox{i}}

\newcommand{\gsim}{\mathrel{\raise.3ex\hbox{$>$\kern-.75em\lower1ex\hbox{$\sim$}}}}
\newcommand{\lsim}{\mathrel{\raise.3ex\hbox{$<$\kern-.75em\lower1ex\hbox{$\sim$}}}}

\begin{document}


\title{ Single-spin asymmetry in DVCS: fragmentation region of
polarised lepton}

\author{I.~Akushevich\thanks{on leave of absence from the National
Center of Particle and High Energy Physics, 220040 Minsk, Belarus}	}
\address{North Carolina Central University,
Durham, NC 27707 and TJNAF, Newport News, VA 23606, USA}
\author{E.A.~Kuraev, B.G.~Shaikhatdenov\thanks{on leave
from the Institute of Physics and Technology, Almaty-82 } }
\address{ JINR, 141980 Dubna, Russia }

\maketitle



\begin{abstract}
For the kinematical region when a hard photon is emitted predominantly
close to the direction of motion of a longitudinally polarized
initial electron and relatively small momentum transfer to a proton
we calculate the azimuthal asymmetry of a photon emission. It arises
from the interference of the Bethe-Heitler amplitude and those which
are described by a heavy photon impact factor.
Azimuthal asymmetry does not decrease in the limit of
infinite cms energy. Lowest order expression for the impact factor
of a heavy photon is presented.
\end{abstract}


\narrowtext

\section{Introduction}
\label{sec:Intro}

Recently a lot of attention has been paid to deeply virtual Compton scattering
(DVCS)~\cite{RAD,XJI,VG,FS,BK}. It was realized that there exists
a close relation with the problem
of a proton spin carried by gluons and quarks. Indeed the decomposition of
the nonforward Compton scattering amplitude for the case when one of photons
on-mass shell and another one is off-mass shell contains fifteen structure
functions and four out of them could be put to the
test~\cite{RAD,XJI,VG,FS,BK}.
Their first moments determine the Dirac, Pauli,
axial-vector and pseudoscalar formfactors of a proton
while their second moments are related with a proton spin fraction
carried out by quarks and gluons and the orbital momenta of the latter.
These structure functions can be tested in DIS
experiments with longitudinally polarized initial lepton
aimed at measuring the azimuthal correlation between
a real photon and scattered lepton.
There are two mechanisms of photon emission, namely
the emission from lepton (Bethe-Heitler) and quark lines.
The last option may be split up into two gauge-invariant mechanisms:
emission off a valence quark of a proton and emission
from a quark-antiquark pair produced by a virtual photon and a gluon
out of a proton.
For the case of small angle scattering (the limit of small Bjorken
variable) the leading nonvanishing in the limit of high cms energies
contribution arises from
the last mechanism that we call the impact factor (IF) mechanism
which is thought of as the amplitude
of virtual photon conversion into real photon in the gluonic field
of a proton.
The effect of azimuthal correlation appears as the interference of the
real Bethe-Heitler amplitude with the pure imaginary one of IF mechanism
of real photon creation.
The interference is not zero due to pure imaginary spin density matrix
of the polarized lepton.

The azimuthal asymmetry we'd like to obtain has the simplest form
${\cal A}=\Delta|M|^2/|M_{BH}|^2 \sim\sin\phi$,
where $\phi$ is the azimuthal angle between the planes formed by the
momenta of initial and scattered leptons and an initial lepton and photon.
It was shown that the higher harmonics in the Fourier decomposition of
the asymmetry are related with the structure functions mentioned above.
The contribution
derived here is sensitive only to the gluon density $zg(z,Q^2)$ inside
a proton.
For small values of energy fraction $z$ carried by sea gluons,
one has $zg(z,Q)\approx 6Q^2\ [GeV^2], Q^2\sim 1\ GeV^2$~\cite{RKZN}.

In what follows we study the case in which an initial proton is unpolarized
and the final states are a scattered lepton, a recoil proton and
a hard photon from the fragmentation region of initial lepton.
Furthermore the lowest order contribution ($\sim\alpha_s^2$) to the
asymmetry is dealt with.
The higher PT effects were considered in the paper~\cite{BK}
and took into account the BFKL ladder.

\section{Bethe--Heitler amplitude}
\label{sec:two}

Let's consider the radiative electron-proton scattering,
$$
e(p_1,\xi) + P(p)\to e(p_2) + P(p') + \gamma(k_1),
$$
where we indicate in parenthesis the 4-momenta of particles,
$\xi$ is the degree of the longitudinal polarization of electron.
We will restrict ourselves to the kinematics where the absolute magnitude
of a square of momentum transfer between initial and final state
electrons is small with respect to the cms energy squared,
\begin{eqnarray}
s=(p_1+p)^2&\gg& Q_1^2=-(p_1-p_2)^2, \\ \nonumber
Q^2=-q^2&\gg& p_1^2=p_2^2=m_e^2, \\ \nonumber
p^2=p^{'2}=M^2,&& q=p'-p.
\end{eqnarray}
The main contribution non-vanishing in the limit of large $s$ arises from
the two Feynman amplitudes. One of them, describing the hard photon emission
by the electron blob, the so called Bethe-Heitler amplitude,
has the following form,
\begin{eqnarray}
M^{BH}_\lambda\!\!\!\! \nonumber \\
&=&\frac{(4\pi\alpha)^{3/2}}{q^2}\bar{u}(p_2)
O_{\mu\sigma} u(p_1,\xi) \bar{u}^\lambda(p')V_\nu u^\lambda(p)
g^{\mu\nu}e^\sigma(k_1) \nonumber \\
&=& \frac{(4\pi\alpha)^{3/2}}{q^2}
\cdot \left(\frac{-2x_1}{sd d_1}\right)\bar{u}(p_2)v_\sigma u(p_1,\xi)
e^\sigma(k_1)s N^\lambda, \nonumber \\
\end{eqnarray}
with
\begin{eqnarray}\label{nla}
V_\nu&=&\gamma_\nu F_1(q^2)+\frac{[\gamma_\nu,\hat{q}]}{2M}F_2(q^2), \\ \nonumber
N^\lambda&=&\frac{1}{s}\bar{u}^\lambda(p')\left(\hat{p}_1 F_1+
\frac{\hat{q}\hat{p}_1}{M}F_2\right) u^\lambda(p), \\ \nonumber
\sum\limits_\lambda|N^\lambda|^2&=&2F(Q^2), \
F(Q^2)=F_1^2(q^2)+\frac{Q^2}{M^2}F_2^2(q^2).
\end{eqnarray}
Here $\lambda=\pm 1$ describes a proton chiral state, $F_{1,2}$
are the Dirac and Pauli form factors, $M$ is a proton mass,
$e(k_1)$ is the photon polarization vector and
$$
v_\sigma=s x(d-d_1)\gamma_\sigma+x d_1\gamma_\sigma\hat{q}\hat{p}
+ d\hat{p}\hat{q}\gamma_\sigma,
$$
the effective vertex describing the Compton scattering~\cite{BFKK}.
The quantities
$$
d=xx_1[(p_1-q)^2-m_e^2],\ d_1=-x_1[(p_1-k_1)^2-m_e^2],\ q^2,
$$
can be re-expressed using the Sudakov decomposition of the 4-vectors,
$$
d=x_1^2m_e^2+(\vecc{k}_1 + x_1\vecc{q})^2,\ d_1=x_1^2m_e^2+\vecc{k}_1^2,
\ Q^2=-q^2=\vecc{q}^2,
$$
where $\vecc{k}_1,\vecc{p}_2,\vecc{q}$ are the two-dimensional
transverse to the beam axis components of photon, scattered electron
and recoil proton momenta which obey the conservation law
$\vecc{k}_1+\vecc{p}_2+\vecc{q}=0$, and $x,x_1$ are the energy
fractions of the scattered electron and real photon satisfying $x+x_1=1$.
The corresponding modulus of the matrix element squared and
summed over polarization states
and the cross section can be brought to the form~\cite{BFKK},
\begin{eqnarray}
\sum|M^{BH}|^2=2^{11}\pi^3\alpha^3\frac{s^2}{\vecc{q}^2}
\frac{x_1^2x(1+x^2)}{dd_1}F(Q^2),
\\ \nonumber
\dd\sigma^{eP\to(e\gamma)P}=\frac{2\alpha^3x_1(1+x^2)}{\pi^2\vecc{q}^2d d_1}
F(Q^2)\dd^2\vecc{k}_1\dd^2\vecc{q} \dd x.
\end{eqnarray}
It is important to note that the amplitude $M^{BH}$ is real.

\section{Asymmetry evaluation}
\label{sec:three}

Consider now the 2-loop level correction to the amplitude studied above,
describing emission of a hard photon from the intermediate state
of a pair of charged quarks created by the virtual photon and converted
to the real one through the two gluons exchange.
The corresponding amplitude differs from the QED one only by the factor
$C=\sum Q_q^2$ ($Q_q$ is a quark charge in units of $e$) and the
gluon density factor $G(z,k,Q)=z\dd g(z,k,Q)/\dd\ln Q^2, k^2\sim Q^2\ll s$
(see Ref.~\cite{RKZN}).
The amplitude of IF mechanism is pure imaginary and
may be expressed in terms of the photon IF,
\begin{eqnarray}
M^{IF}&=&4C\alpha_s^2\alpha\frac{\ii s(4\pi\alpha)^{1/2}}{q_1^2}\bar{u}(p_2)\gamma_{\mu}
u(p_1,\xi)\frac{ N^\lambda}{(2\pi)^2} \\ \nonumber
&\times&\int\frac{\dd^2\vecc{k} G(z,k,Q)}{\pi\vecc{k}^2(\vecc{q}-\vecc{k})^2}
\frac{\dd^2\vecc{q}_+ \dd x_+ }{\pi x_+x_-}I_{\mu\sigma}
e^\sigma(k_1),
\ q_1^2=-\frac{\vecc{p}_2^2}{x},
\end{eqnarray}
where the tensor $I_{\mu\sigma}$ is given through the tensor of elastic
gluon-photon scattering,
\begin{equation}
I_{\mu\sigma}=\frac{p^\alpha p^\beta}{s^2} T_{\alpha\beta\mu\sigma}.
\end{equation}
Its explicit form is given in Appendices A and C. In the last appendix
we infer the heavy photon IF with both photons off-mass shell.

The relevant expression for the contribution to the cross section reads,
\begin{eqnarray}
\Delta|M|^2&=&\sum 2M^{IF}(M^{BH})^* \nonumber \\
&=&s^2\xi 2^{11}C\frac{x x_1\pi^2}{\vecc{q}^2\vecc{p}_2^2 d d_1}
\alpha^3\alpha_s^2\int\frac{\dd^2\vecc{k} G(k,Q,z)}
{\pi\vecc{k}^2(\vecc{q}-\vecc{k})^2} \nonumber \\
&\times&\int\frac{\dd^2\vecc{q}_+ \dd x_+ }{x_+x_-\pi}\cdot
J\cdot F_1(Q^2), \\ \nonumber
J&=&\frac{\ii}{s} I_{\mu\nu}L_{\mu\nu},\quad
L_{\mu\nu}=\frac{1}{4}{\mathrm{Tr}}[\hat{p}_2v_\nu\hat{p}_1\gamma_\mu\gamma_5].
\end{eqnarray}
Using the gauge invariance conditions
$T_{\alpha\beta..}k_\alpha=T_{\alpha\beta..}(q-k)_\beta=0$
we can perform the following replacement in the expression
for $I_{\mu\sigma}$,
\begin{eqnarray}
\frac{p_\alpha p_\beta}{s^2}&\to&
\frac{k^\bot_\alpha(q-k)^\bot_\beta}{\tilde ss_1'}, \\ \nonumber
\tilde s=\frac{1}{x_1}[(q_++q_-)^2&+&(\vecc{q}_1+\vecc{k})^2],\\ \nonumber
s_1'=\frac{1}{x_1}[(q_++q_-)^2&+&(\vecc{k}_1+\vecc{k}-\vecc{q})^2].
\end{eqnarray}
The next step is to perform the $\dd^2\vecc{k}$ integration. We
suppose that small values of $|\vecc{k}|$ dominate as this region
is enhanced by the factor $zg(z,|\vecc{k}|)$.
Then the integration could be carried out as follows,
\begin{eqnarray}
&&\int\frac{\dd^2 \vecc{k}}{\pi\vecc{k}^2\vecc{k}^{'2}}
\vecc{k}^i \vecc{k}^{'j}G(z,\vecc{k},\vecc{k}')
\nonumber \\ &=&\int\limits_0^1\dd x
\int\frac{\dd^2 \vecc{k}\vecc{k}^i(\vecc{q}-\vecc{k})^j
G(z,\vecc{k},\vecc{q}-\vecc{k})}
{\pi[(\vecc{k}-x\vecc{q})^2+Q^2x(1-x)]^2}		\\ \nonumber
&\approx& \frac{1}{2}\delta^{ij}\int\limits_0^1 \dd x zg(z,xQ,(1-x)Q)
\approx \frac{1}{2}\delta^{ij} zg(z,Q/2).
\end{eqnarray}
Thus to the accuracy of approximately $10$\% (with $Q^2$ of a few $GeV^2$) 
we may put $|\vecc{k}|=|\vecc{k}'|=Q/2$
in the nonsingular part of the integrand.

It should be noted that only the structure
$$
E=(p,p_1,p_2,q)=\varepsilon_{\alpha\beta\gamma\delta}
p^\alpha p_1^\beta p_2^\gamma q^\delta
=\frac{s}{2}\left[\vecc{p}_2\times\vecc{q}\right]_z
$$
survives integrations over
$\dd^2\vecc{q}_+,\dd x_+$ (for details see appendices A,B).
Then the asymmetry is found to be,
\begin{eqnarray}\label{main}
&&{\cal A}=\xi\frac{\alpha_s^2}{\pi}\frac{zg(z,Q/2)|_{z\to 0}F_1(Q^2)}{F(Q^2)}
\left|\frac{\vecc{q}}{\vecc{p}_2}\right|\Phi(x)\sin\phi,  \\ \nonumber
&&\Phi(x)=-\frac{1+x}{3[1+x^2]}
\left(2\ln\frac{\vecc{p}_2^2}{m^2} -1\right),
\end{eqnarray}
Here $Q=|\vecc{q}|$ is the momentum transfer to a proton
which bears to a some extent a latent dependence on an energy 
fraction of the scattered lepton,
$|\vecc{p}_2|$ is
the transverse component of the scattered electron momentum,
$m=0.3\ GeV$ is a quark constituent mass.
Besides it has been assumed that $\vecc{p}_2^2 \gg Q^2\gg m^2$
and the terms of order $(m^2/\vecc{p}^2_2)\ln(\vecc{p}^2_2/m^2)$
have been dropped for their subleading nature (which gives an
accuracy of the derivation to be $\sim 10\%$)\footnote{Note
that we use the notation typical for calculations based on Sudakov
parameterization of four-vectors. The invariants do not always
coincide with those used by experimentalists. 
Thus our $x,\ Q^2_1=\vecc{p}_2^2/x$ and
$Q^2$ correspond to $1-y$, $Q^2$ and $-t$, in order, adopted 
by experimental collaborations; $y$ is the scaling variable.}.

\section{Conclusion}
\label{sec:concl}
Above we gave a rather rough estimate for the azimuthal asymmetry
of a real photon emission induced by longitudinally polarized electron
in its fragmentation region. It turns out that the asymmetry is enhanced
by a sea gluon density $zg(z,Q)$ which for $z\to 0$ appears to be
$\sim 5\div 7\ (Q(GeV))^2$.
Evidently the asymmetry results from the interference between
the born-level Bethe-Heitler amplitude and that of two-loop level
containing a photon-gluon fusion block. The first amplitude is real
and the last one is completely imaginary.
Aiming at obtaining a definite analytical result for the asymmetry
we study the problem restricted by the requirements
$\vecc{p}_2^2\gg Q^2\gg m^2$. Using this approximation we extract
the gluon density factor $zg(z,Q)$. The non-enhanced terms are estimated
to give a contribution of order unity thereby claiming the accuracy
of the calculation to be of order $15\div 20\%$.
Just to illustrate what the asymmetry looks like we give a plot
of the beam-spin analyzing power values (which is merely a multiplicative
factor in front of $\sin\phi$ in Eq.~(\ref{main}))
for the momentum transfer $Q_1^2=1\div 50$~GeV$^2$. The
ratio of $|\vecc{q}/\vecc{p}_2|$ is kept to be 0.1 for all curves 
in the plot. The fall of the asymmetry
for high values of $x$ and $Q_1^2$ is caused by relatively
large values of  $|\vecc{q}|$ and, as a result, by the little bit less 
$\alpha_s$ values.

\begin{figure}[!ltb]
\begin{picture}(80,80)
\put(0,5){
\epsfxsize=8cm
\epsfysize=8cm
\epsfbox{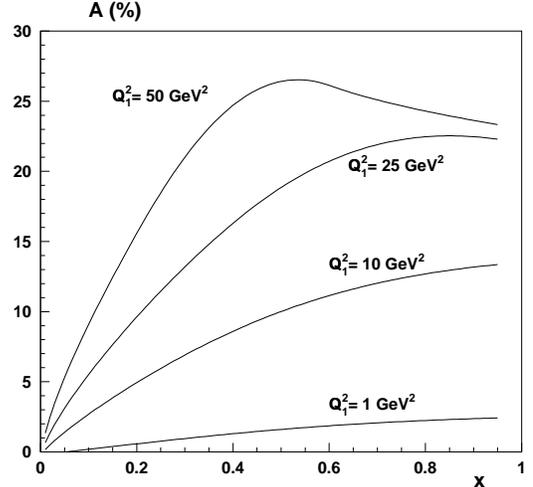}
}
\end{picture}
\vskip-1.2cm
\caption{ The beam-spin analyzing power
${\bf A}(x)$ versus $x$ for
$Q^2_1= 1\div 50$ GeV$^2$ and $|\vecc{q}/\vecc{p}_2|$=0.1.
}
\end{figure}

\vskip 1cm
{\bf Acknowledgements}:
The authors are grateful to B.V.~Struminsky for useful discussions.
The work of IA was supported by the
U.S. Department of Energy under contract DE--AC05--84ER40150 while
that of EAK and BGS --- by the RFBR 99-02-17730 and HLP 2001-02 grants.

\section*{Appendix~A: Explicit form of $I_{\mu\nu}$}
\label{sec:appena}
\setcounter{equation}{0}
\renewcommand{\theequation}{A.\arabic{equation}}

To calculate the contribution of FD containing the impact
factor of a heavy photon we need the trace,
\begin{eqnarray}\label{a1}
I_{\mu\nu}&=&\frac{1}{s^2}{\mathrm{Tr}}
[(\hat q_- + m)B_\mu (\hat q_+ - m)R_\nu], \\ \nonumber
B_\mu&=&\frac{1}{d_+}\gamma_\mu(\hat k - \hat q_+ + m)\hat p_2
+ \frac{1}{d_-}\hat p_2(\hat q_- - \hat k + m)\gamma_\mu, \\ \nonumber
R_\nu&=&\frac{1}{d_-'}\gamma_\nu(\hat q_- - \hat k' + m)\hat p_2
+ \frac{1}{d_+'}\hat p_2(\hat k' - \hat q_+ + m)\gamma_\nu,
\end{eqnarray}
where $m$ is a quark mass and
$$ d_{\pm}=k^2-2kq_{\pm},\qquad d_{\pm}'=k^{'2}-2k'q_{\pm}.$$
It's easy to show that the gauge conditions for the on-mass shell quarks
are satisfied,
\begin{equation}
\bar{u}(q_-)B_\nu v(q_+)q_1^\nu=0,\quad\bar{v}(q_+)R_\mu u(q_-) k_1^\mu=0.
\end{equation}

Taking into account an enhancement due to large gluon
density factor $zg(z,Q)\gg 1$ one can restrict further consideration
to the kinematics $\vecc{p}_2^2\gg\vecc{k}^2\sim\vecc{q}^2$
which is thus preferable. Then it could be verified that
\begin{eqnarray}
d_{\pm}&=&d_{\pm}'=-x_{\pm}\tilde s,\quad
\tilde s = \frac{1}{x_1}\left[s_1+\vecc{q}_1^2\right]
=\frac{\sigma}{x_1y_+y_-},  \nonumber \\
s_1&=&(q_++q_-)^2=\frac{1}{y_+y_-}[m^2+\vecc{q}_t^2], \\ \nonumber
\sigma&=&m_*^2+\vecc{q}_t^2, \qquad m_*^2=m^2+y_+y_-\vecc{p}_2^2, \\ \nonumber
\vecc{q}_t&=&\vecc{q}_+ +y_+\vecc{p}_2=-\vecc{q}_- - y_-\vecc{p}_2,\quad
y_{\pm}=\frac{x_{\pm}}{x_1}.
\end{eqnarray}
Here $x_{\pm}$ are the energy fractions of the pair,
$\vecc{q}_{\pm}$ --- their components of momentum transverse
to the beam axis. They obey the conservation laws,
$$
y_++y_-=1,\quad{\mathrm{and}}\quad\vecc{q}_+ +\vecc{q}_- + \vecc{p}_2=0.
$$

With the substitution $p_{2\mu}\to -sk_\mu^\bot/\tilde{s}$
in the quantity $B$ and, respectively,
$p_{2\mu}\to -sk_\mu^{'\bot}/\tilde{s}$ in $R$
the tensor $I_{\mu\nu}$ can be transformed to take the following form,
\begin{eqnarray}
I_{\mu\nu}&=&\frac{1}{4\tilde s^4}{\mathrm{Tr}}
[(\hat q_-+m)B_{1\mu}(\hat q_+ - m)R_{1\nu}], \\ \nonumber
B_{1\mu}&=&\frac{\tilde s}{s}\left(\frac{1}{x_+}\gamma_\mu \hat p \hat k
-\frac{1}{x_-}\hat k \hat p\gamma_\mu\right) + \gamma_\mu Z, \\ \nonumber
R_{1\nu}&=&\frac{\tilde s}{s}\left(-\frac{1}{x_-}\gamma_\nu \hat p \hat k'
+\frac{1}{x_+}\hat k' \hat p\gamma_\nu\right) + \gamma_\nu Z', \\ \nonumber
Z&=&\frac{2}{x_+x_-}\vecc{r}\vecc{k},\quad
Z'=\frac{2}{x_+x_-}\vecc{r}\vecc{k}',\quad \vecc{r}=x_1\vecc{q}_t.
\end{eqnarray}
Here the vectors $\vecc{k},\ \vecc{k}'=\vecc{k}-\vecc{q}$
are pure 2-dimensional ones transverse to the beam axis.

Once again one can check that the gauge conditions,
\begin{eqnarray}
&&\bar{u}(q_-)B_{1\nu} v(q_+)q_1^\nu=0, \nonumber \\  \\ \nonumber
&&\bar{v}(q_+)R_{1\mu} u(q_-) k_1^\mu=0,
\end{eqnarray}
are satisfied up to the terms of order $\vecc{k}^2/\vecc{p}_2^2$.

\section*{Appendix~B: Integration over quark pair momenta}
\label{sec:appenb}
\setcounter{equation}{0}
\renewcommand{\theequation}{B.\arabic{equation}}

Using the Sudakov parametrization of 4-vectors,
\begin{eqnarray}
&&p_1\approx\tilde p_{1},\quad p_2=x\tilde p_{1}
+\frac{\vecc{p}_2^2}{xs}\tilde p + p_{2\bot},  \nonumber \\
&&q_{\pm}=x_{\pm}\tilde p_{1}+\frac{\vecc{q}_{\pm}^2+m^2}
{sx_\pm}\tilde p + q_{\pm\bot}, \\ \nonumber
&&k_1=x_1\tilde p_{1} + \frac{\vecc{k}_1^2}{sx_1}\tilde p + k_{1\bot},\quad
\tilde p_1^2=\tilde p^2=0, \\ \nonumber
&&\tilde p=p-\frac{M^2}{s}p_1,\quad 2\tilde p\tilde p_1=s,
\quad a_\bot \tilde p_1=a_\bot \tilde p=0\,,
\end{eqnarray}
and the conservation law $x_+ + x_-=x_1$,
the scalar products used can be written as follows,
\begin{eqnarray}
2p_1q_{\pm}&=&\frac{1}{x_{\pm}}[\vecc{q}_{\pm}^2+m^2]\equiv a_{\pm}, \nonumber \\
2p_2q_{\pm}&=&\frac{1}{xx_{\pm}}[m^2x^2
+(x\vecc{q}_{\pm}-\vecc{p}_2x_{\pm})^2]\equiv a'_{\pm}.
\end{eqnarray}
Upon averaging over the azimuthal angle of gluon momenta
and using the permutation symmetry
\begin{eqnarray*}
x_-\longleftrightarrow x_+\, ,\qquad
\vecc{q}_-\longleftrightarrow \vecc{q}_+\, ,
\end{eqnarray*}
the quantity $J$ could be symbolically written in the following manner,
\begin{eqnarray}\label{b3}
\frac{s\tilde s^4}{\vecc{p}_2^2}\!\!&\cdot&\!\!\frac{J}{\vecc{k}^2}=
\frac{1}{2}(1+{\cal P}_\pm)[xB+C],\\ \nonumber
C&=&E\left(\frac{2s_t^2}{x_1^2}-\frac{4m^2\vecc{r}^2}{z^2}\right)
+ 2(p,p_2,q_-,q)\biggl(-\frac{2s_t}{x_1x_+z}\vecc{q}_+\vecc{r}
\\ \nonumber
&+&\frac{s_t^2}{x_1z}+\frac{2a_+}{z^2}\vecc{r}^2-\frac{s_t^2}{x_-x_1^2}\biggr)
-2(p,p_2,q,r)\frac{s_ta_+}{x_1z},
\\ \nonumber
B&=&2(p,p_1,q_-,q)\Biggl(-\frac{s_t}{x_1z}\vecc{r}\vecc{p}_2
+\frac{xs_t}{x_+x_1z}\vecc{r}\vecc{q}_+ - \frac{xs_t^2}{2x_1z} \\ \nonumber
&-&\frac{\vecc{r}^2a_+'}{z^2}
+\frac{xs_t^2}{2x_-x_1^2}\Biggr) + (p,p_1,q,r)\frac{s_ta_+'}{x_1z}
+(p,p_1,p_2,r) \\ \nonumber
&\times&\frac{2s_t}{x_1z}\vecc{q}_-\vecc{q}
+2(p,p_1,p_2,q_-)\left(\frac{s_t}{x_1z}\vecc{r}\vecc{q}
-\frac{2}{z^2}\vecc{r}^2\vecc{q}_+\vecc{q}\right)  \\ \nonumber
&+&E\left[-\frac{2s_t(x_1+x_-)}{x_1x_-z}\vecc{r}\vecc{q}_-
-\frac{s_t^2}{2z}-\frac{2}{z^2}\vecc{r}^2(s_t-\vecc{p}_2^2)\right]
\\ \nonumber
&+&2(p,p_2,q_-,q)\biggl(\frac{s_t}{x_1x_+z}\vecc{q}_+\vecc{r}
-\frac{s_t^2}{2x_1z} - \frac{a_+}{z^2}\vecc{r}^2  \\ \nonumber
&+& \frac{s_t^2}{2x_-x_1^2}\biggr) +(p,p_2,q,r)\frac{s_ta_+}{x_1z}
+s\biggl[2(p_1,p_2,q_-,q) \\ \nonumber
&\times&\frac{x_+\vecc{r}^2}{z^2}
+ (p_1,p_2,q,r)\frac{x_-s_t}{x_1z}\biggr],
\end{eqnarray}
where $s_t=x_1\tilde s$ and $z=x_+x_-$ and ${\cal P}_\pm$ is the
permutation operator.
In deriving these formul\ae{} it has been assumed that
$\vecc{k}^2\gg\vecc{q}^2$.
The structures $(\ldots)$ entering Eq.~(\ref{b3}) can be rewritten
as follows\footnote{At this point one should be aware that only the
transverse components of the 4-vector q have to be taken into account.},
\begin{eqnarray*}
&(p,p_2,q_-,q)=x(p,p_1,q_-,q) - x_-E,& \\
&s(p_1,p_2,q_-,q)=-\displaystyle{\frac{\vecc{p}_2^2}{x}}(p,p_1,q_-,q)
+ a_-E,& \\
&s(p_1,p_2,q,r)=\displaystyle{\frac{\vecc{p}_2^2}{x}}(p,p_1,r,q),& \\
&(p,p_2,q,r)=-x(p,p_1,r,q).&
\end{eqnarray*}
Having all the above at hand we turn to the $\dd^2\vecc{q}_+$
integration. A set of relevant integrals reads,
\begin{eqnarray}
&&\int\frac{\dd^2\vecc{q}_+}{\pi}\Biggl\{\frac{1}{\sigma^2},
\frac{\vecc{r}^2}{\sigma^4}, \frac{\vecc{q}_-^i\vecc{r}^j}{\sigma^3},
\frac{\vecc{r}^i}{\sigma^3}, \frac{\vecc{r}^2\vecc{q}_+^i\vecc{q}_-^j}
{\sigma^4}, \frac{(\vecc{r}\vecc{q_+})\vecc{q}_-^i}{\sigma^3},	\nonumber \\
&&\frac{\vecc{r}^2a_+'\vecc{q}_-^i}{\sigma^4},
\frac{\vecc{r}^2a_+\vecc{q}_-^i}{\sigma^4}, \frac{a_+'\vecc{r}_i}{\sigma^3},
\frac{a_+\vecc{r}_i}{\sigma^3}, \frac{\vecc{r}^2a_{\pm}}{\sigma^4},
\frac{a_{\pm}}{\sigma^3}, \frac{\vecc{q}_+\vecc{r}}{\sigma^3}
\Biggr\}  \nonumber \\
&=&\Biggl\{\frac{1}{m_*^2},\frac{x_1^2}{6m_*^4},
-\delta^{ij}\frac{x_1}{4m_*^2}, 0,
-x_1^2\left(\frac{\delta_{ij}}{6m_*^2}-\vecc{p}_2^i\vecc{p}_2^j\frac{y_+y_-}
{6m_*^4}\right), \nonumber \\
&&\frac{x_1\vecc{p}_2^i}{2m_*^2}\left(\frac{1}{2}y_+ - y_-\right),
\frac{x_1^2x\vecc{p}_2^i}{3m_*^2x_+}\left(\frac{y_+}{x} - y_-
- \frac{y_-y^2_+\vecc{p}^2_2}{2x^2m^2_*}\right),  \nonumber \\
&&\frac{x_1^2\vecc{p}_2^i}{3m_*^2x_+}
\left(y_+ - y_- - \frac{y_-y^2_+\vecc{p}^2_2}{2m^2_*}
\right),  -\frac{\vecc{p}_2^i}{2m_*^2},
- \frac{\vecc{p}_2^i}{2m_*^2}, \\ \nonumber
&&\frac{x_1^2}{3x_{\pm}m_*^2}
\left(1+\frac{y_{\pm}^2\vecc{p}_2^2}{2m_*^2}\right),
\frac{1}{2x_\pm m_*^2}\left(1+\frac{y_{\pm}^2\vecc{p}_2^2}{m_*^2}\right),
\frac{x_1}{2m_*^2}\Biggr\}.
\end{eqnarray}
In the above we have discarded the terms that give the contributions
of order $m^2/\vecc{p}_2^2$ as compared with unity.
The integration over $y_{\pm}$ becomes almost trivial in the limit
$Q_1^2\gg m^2$,
\begin{eqnarray}
\int\limits_0^{1}\frac{\dd y_ +}{m_*^2}\biggl\{1,y_{\pm},y_{\pm}^2,
\!&&\!\!\!y_+y_-,\frac{m^2}{m^2_*}y_-y_+\biggr\} \\ \nonumber
&=& \frac{1}{\vecc{p}_2^2}\{2L,L,L-1,1,0\},
\end{eqnarray}
where $L=\ln(\vecc{p}^2_2/m^2)$.

\section*{Appendix~C: Heavy photon impact factor}
\label{sec:appenc}
\setcounter{equation}{0}
\renewcommand{\theequation}{C.\arabic{equation}}

To obtain the heavy photon IF one has to consider the s-channel discontinuity
of the heavy photon amplitude in an external field,
\begin{eqnarray}
\gamma_\mu(P_1) + A(p)&\to& q(q_+) + \bar{q}(q_-) + A(p^{''}) \nonumber \\
&\to& \gamma_\nu(P_2) + A(p'), \\ \nonumber
P_1^2=-Q^2,&& P_2^2=-Q^{'2},
\end{eqnarray}
which is described by the tensor
\begin{equation}
\Delta A_{\mu\nu}(s,t)=\frac{(4\pi\alpha)^3}{\vecc{k}^2\vecc{k}^{'2}}
\left(\frac{2}{s}\right)^2N_\lambda s^4 I_{\mu\nu}\dd\Gamma_3,
\end{equation}
with
\begin{eqnarray}
\dd\Gamma_3&=&\frac{1}{(2\pi)^5}\frac{\dd^3\vec{q}_+}{2\varepsilon_+}
\frac{\dd^3\vec{q}_-}{2\varepsilon_-}\frac{\dd^3\vec{p}^{''}}{2E^{''}}
\delta^4(P_1+p-q_+-q_--p^{''}) \nonumber \\
&=&\frac{\dd^2\vecc{k}\dd^2\vecc{q}_+\dd x_+}{4s(2\pi)^5x_+x_-},
\end{eqnarray}
and the quantity $N_\lambda$ given in the Eq.~(\ref{nla}).
The tensor $I_{\mu\nu}$ has the form (see Eq.~(\ref{a1})),
\begin{eqnarray}
\frac{1}{x_+x_-}I_{\mu\nu}&=&(1+{\cal P}_\pm)\Biggl\{
\frac{x_+}{a_+a'_+}\bigl(2q_\mu q_{+\nu} + (2-4x_+)q_\nu q_{+\mu} \nonumber \\
&-& 8x_-q_{+\mu}q_{+\nu}\bigr) + \frac{1}{a_-a_+'}\bigl(2x_+q_\mu q_{-\nu}
+ (-2x_+  \nonumber \\
&+&4x_+x_-)q_\nu q_{-\mu} - 2q_{-\nu} q_{+\mu} + (2-8x_+x_-)  \nonumber \\
&\times&q_{+\nu}q_{-\mu}\bigr)
+ g_{\mu\nu}\biggl[\frac{1}{a_-a_+'} \bigl(-x_-\vecc{k}^2
-x_+\vecc{k}^{'2}  \nonumber \\
&+& x_+x_-(\vecc{q}^2-Q^2-Q^{'2}) \bigr)
+ \frac{x_+}{a_+a_+'}(x_-(Q^2  \nonumber \\
&+&Q^{'2})+ x_+\vecc{q}^2)\biggr]\Biggr\},
\end{eqnarray}
with
\begin{eqnarray*}
a_{\pm}&=&a+\vecc{q}_{\pm}^2,\quad
a_{\pm}'=b+(\vecc{q}_{\pm}-x_{\pm}\vecc{q})^2, \\
a&=&m^2+x_+x_-Q^2,\quad b=m^2+x_+x_-Q^{'2}.
\end{eqnarray*}
One can argue that the gauge condition $I_{\mu\nu}=0$
for $k=0, k'=0$ is satisfied. Joining the denominators with the use of
the Feynman trick and performing an integration over the transverse
to the beam axis components of the quark pair momenta we get,
\begin{eqnarray}\label{c5}
\int\frac{\dd^2\vecc{q}_+}{\pi a_+a_+'}&=&\int\limits_0^1\frac{\dd y}{D_{++}},
\qquad
\int\frac{\dd^2\vecc{q}_+}{\pi a_+a_-'}=\int\limits_0^1\frac{\dd y}{D_{-+}},
	\\ \nonumber
D_{++}&=&A+\vecc{q}^2x_+^2y(1-y), \\ \nonumber
D_{-+}&=&A+y(1-y)\vecc{b}^2, \\ \nonumber
\vecc{b}&=&\vecc{k}-x_+\vecc{q}, \\ \nonumber
A&=&m^2+x_+x_-(yQ^{'2} + (1-y)Q^2).
\end{eqnarray}
The result for IF takes the following form
(we choose only the transverse polarizations of photons $\mu\equiv i,
\nu\equiv j$):
\begin{eqnarray}
\tau^\gamma_{ij}&=&2\alpha^2\int\limits_0^1\dd x_+\dd x_-\delta(x_++x_--1)
\int\limits_0^1\dd y\biggl[\frac{x_+^2}{D_{++}} \nonumber \\
&\times&[8x_+x_-y(1-y)q_iq_j-\vecc{q}^2\delta_{ij}(1+
4x_+x_-y(1-2y))]  \nonumber \\
&-&\frac{1}{D_{-+}}\biggl[8x_+x_-y(1-y)b_ib_j-\vecc{b}^2\delta_{ij}
(1 + 4x_+x_-y ( 1 \nonumber \\
&-&2y))\biggr] + 4x_+x_-y(x_--x_+)q_j\left(\frac{x_+q_i}{D_{++}}
+\frac{b_i}{D_{-+}} \right)  \nonumber \\
&+&\delta_{ij}x_+x_-(Q^2+Q^{'2}+4x_+x_-y(Q^{'2}-Q^2)) \nonumber \\
&\times&\left(\frac{1}{D_{-+}}-\frac{1}{D_{++}}\right)\biggr],
\end{eqnarray}
with $D_{++}, D_{-+}$ and $\vecc{b}$ defined the same way as
in the Eq.~(\ref{c5}).
Once again it is clearly seen that the gauge conditions are satisfied,
$$
\tau|_{\vecc{k}=0}=\tau|_{\vecc{k}=\vecc{q}}=0.
$$
It is important to note that even for the on-mass shell photons
$Q^2=Q^{'2}=0$
this expression differs from the one derived by Cheng and Wu~\cite{CWGLF}.
The difference is found to be
\begin{eqnarray}
\Delta \tau^\gamma_{ij}&=&\tau-\tau_{CW} \\  \nonumber
&=&4\alpha^2\int \dd x_+\dd y\ x_+x_-(1\!\!-\!\!2x_+)q_j
\left(\frac{x_+q_i}{D^0_{++}} + \frac{b_i}{D^0_{-+}}\right), \\ \nonumber
D^0_{++}&=&m^2+x_+^2y(1-y)\vecc{q}^2,\
D^0_{-+}=m^2 + y(1-y)\vecc{b}^2.
\end{eqnarray}
The reason for this discrepancy is in the different definition of
initial and final photons' 4-momenta. The similar results were obtained
in Ref.~\cite{Dav}.

\end{document}